\documentclass[a4paper,12pt]{article}
\usepackage[a4paper,text={16.8cm,22.4cm}]{geometry}
\usepackage{amsmath,amssymb,latexsym,bm}
\usepackage{cite}

\def\la{\mathrel{\mathpalette\fun <}}

\def\fun#1#2{\lower3.6pt\vbox{\baselineskip0pt\lineskip.9pt
\ialign{$\mathsurround=0pt#1\hfil##\hfil$\crcr#2\crcr\sim\crcr}}}

\begin{document}

\begin{titlepage}

\begin{center}
  \vspace{5ex}
\textbf{ \Large Nonstandard  Origin of the  Standard  Electroweak
Currents} \vspace{7ex}

\textsc{I. T. Dyatlov} \vspace{2ex}

\textsl{ Petersburg Nuclear Physics Institute, Gatchina
188350,Russia }

\vspace{4ex}
\begin{abstract}\noindent
Implications are considered of the hypothesis that the symplectic
group $Sp(n/2)$ is the spontaneously violated gauge group of n
lepton flavors. Invariant Majorana masses are impossible in
$Sp(n/2)$. For the local gauge symmetry $Sp(n/2)$ the dynamical
spontaneous violation  is only achievable for the number of
flavors $n=6$ with simultaneous parity ($R,L-$symmetry) violation.
The see-saw mechanism produces here three light and three heavy
Dirac neutrinos. Majorana states are unavailable here. Neglecting
heavy particles in the $R,L-$symmetric system of weak and
electromagnetic interactions ($R,L-$independent values of isospins
Tw and  hypercharges $Y$ for leptons or quarks ) leads to a theory
with parity nonconservation and axial anomalies. Only weak left
$(L)$ and full $(R+L)$ electromagnetic currents do not have
anomalies and remain independent of the physics of heavy masses.
These currents are the  ones of the Standard Model. The absence of
anomalies merely in the combination of currents forming the
electromagnetic one presents essential difference with the SM
case, where the both, left
      $T_W$
and $Y$, currents are deprived of anomalies independently.

      PACS        numbers:  12.60,-i, 12.60.NZ, 11.15,-q
\end{abstract}

\end{center}

\end{titlepage}


\section {Introduction}

Neutrinos are unique particles. Firstly, they participate only in weak
interactions (of those known to us). Although neutrinos have a mass,
as any other particles, they are the only ones whose dynamics are
characterized by chirality.
One can easily assume that weak interactions
themselves are determined by the properties of neutrinos $(\nu)$,
whereas charged leptons and quarks adjust themselves to the structure
of $\nu$ states.

This view is discussed in the present paper and is different from
the conventional approach to the electroweak part of the Standard
Model (SM) \cite{d1} and the theory of mass spectrum and neutrino
mixing (see \cite{d2} and also reviews \cite{d3,d4} with numerous
references). The differences of our approach are discussed later
in this section.

      Secondly, only neutrinos can have the Majorana
form.  This form is equivalent to chiral states: for massless
particles, two states of the Majorana particle spin precisely
correspond to the particle-antiparticle pair in the chiral
representation (for right $(R)$ and left $(L)$). It is therefore the
Majorana form that can play a major role in the formation of properties
of both neutrinos and weak interactions.

Thirdly, the $\nu$ mass spectrum \cite{d2} is apparently based on
other
     principles than masses of charged and therefore compulsorily Dirac
     particles. The exceptional smallness of masses, the absence of any
     visible hierarchy of generations, and a very high value of the
     ratio between mass-squared differences suggest essentially
     different dynamics of neutrino mass formation.
     Using standard  conventions \cite{d2}, we have:
     \begin{equation}
23\la \Delta m_{23}^2/\Delta m_{12}^2\la 43
\end{equation}
     The second and third points listed above were discussed and proved
     to be correct for the mechanism of spontaneous $\nu$ mass
     formation based on the gauge coupling of $n$ flavors in the
     symplectic group $Sp(n/2)$ \cite{d5}.

The $Sp(n/2)$ group is distinguished
     because the invariant Majorana mass, both for right $R$ (chiral)
     and left $L$ states, is identically equal here to zero. This
     property is demonstrated only by the fundamental spinor
     representation of $n$ in the $Sp(n/2)$ group. Any other
     representations and groups are devoid of it. Therefore, the
     generation of Majorana masses in $Sp(n/2)$ is only possible upon
     total destruction of the group: the n mass matrix may appear at
     once and individual diagonal and nondiagonal elements may be
     different from zero. This results in a whole spectrum of masses
     different in flavor, rather than one mass similar for all flavors.
     At the same time, the Dirac part of the general n mass matrix may
     be $Sp(n/2)$-invariant and Dirac masses may be the same for all
     flavors.

Conventionally, symmetry violation: standard
     electroweak symmetry, grand unification symmetry, or generation
     flavor symmetry (\cite{d3,d4} and report \cite{d6}),   is realized
     through vacuum averages of the various Higgs scalar fields. At
     that, parity nonconservation and the number of generations are
     phenomenologically postulated.

     The $Sp(n/2)$ group properties
      become of interest if there is a possibility of non-Higgs,
      dynamical symmetry violation. Assuming the violation of this kind
      is possible, one may gain an insight into those SM aspects
      (parity violation dynamics and the number of generations, among
      others) that cannot be explained by conventional approaches.
      However, the nonperturbative hypothesis and the absence of
      quantitative solutions certainly reduce the significance of the
      results.

     In this connection, let us note two restrictions that
     are imposed on this problem:

     1. The dynamics of real gauge
theories are too complicated. Therefore, we consider only possible
symmetry consequences of the hypothesis on the violated $Sp(n/2)$
symmetry of lepton flavors.

     2. We assume that symmetry consequences
and properties of mass equations (major elements in the process of
mass generation) can be observed if a significant part of the
gauge theory dynamics is neglected: for example, by using
Nambu-Jona-Lasinio fermion propagators \cite{d7}: $S_F^{-1}(p) =
U^+(\hat p-M_{diag})U$, where $U$ is a diagonalizing matrix. At
the same time, the conclusion on the number of neutrino flavors
(see below) seems to be more general.

     Tuning conditions merely to sustain the system of equations
("gap equations" \cite{d7}) for the unambiguous determination of
parameters of this highly complex, spontaneously appearing mass
matrix ($2n\times 2n$, $L+R$ neutrinos) is difficult to achieve
and imposes strict constraints on the system and the matrix
itself. Tuning is only possible when considering the n-system as a
quasi-Majorana form, where $Sp(n/2)-$covariant superpositions of
chiral neutrinos-antineutrinos (analogs of Bogolyubov
"particle-hole" states \cite{d8}) are assumed to define the
development of process dynamics.

The physically interesting
     scenario for the local gauge $Sp(n/2)$ theory requires the
     following compulsory conditions \cite{d5}:

      a) $n=6$ for the number of n flavors,

      b) the Majorana mass matrix $(M_{LL})$ of $\nu$ left
     flavors ($L\rightleftharpoons R$ is certainly possible here) vanishes.

     Item
(b) means the breaking of the $R,L-$symmetry, or parity.
Simultaneously, this condition (b) is a requirement of implementation
of one of the two necessary elements that engage the see-saw mechanism
     (see review \cite{d4}).

      Then, the suggestion that the Majorana
     scale $M$ is much greater than the Dirac scale $\mu$ (the second
     condition of the see-saw), brings us $(Sp(3)$ to a system of
     three light $(\sim\mu^2/M)$  and three heavy $(\sim M)$
neutrinos in $Sp(n/2)$ \footnote{Note that due to the smalines of
neutrino masses, the parameter $M$ is very large: for $\mu\sim\mu_\tau
\sim 1$ GeV $-M\simeq 10^{11}$ GeV, for $\mu\sim\mu_W\sim 10^2$ GeV -
$M\simeq 10^{15}$ GeV}.
 All $\nu$ appear to be Dirac, as the characteristic equation for
the Majorana spectrum resolves into mutually similar (in absolute
value) eigenvalues \cite{d5}.

      Relation (1) is easily reproduced by
     the appearing matrices. The great value of (1) does not mean,
     though, any degeneration of states (1,2), as it does in the usual
      Higgs system where $\Delta m^2_{12}$ is associated with the
      exclusively small difference of Dirac mass $\mu_{ik}$
      eigenvalues.  In the $Sp(3)$ system, values (1) are determined by
      the difference of the Majorana parameters $M_i$ that, being
      considerably different from each other
      $(m_{\nu_i}\sim\mu^2/M_i)$, can reproduce (1).

If we employ several types of scalar fields with various $Sp(n/2)$
characteristics ( including the $Sp(n/2)$ adjoint representation,
satisfying eq. (8) ) and weak isospins, it would be
 possible to construct lepton mass matrices at any $n$. But such
      a mechanism (similar to the usual Higgs one) hardly
      compatible with the weak properties of leptons : Majorana
      and Dirac masses of neutrinos, necessarily Dirac- type
      masses for charged leptons, various weak properties of
      $R$ and $L$ fermion states.  At the same time, the dynamical
      spontaneous violation  by the use of quasi-Majorana lepton states
      helps simultaneously to determine the up and down weak
      components, to note a difference of its $R,L-$properties and is
      able to produce a proper neutrino spectrum. These were guides for
      a choice of the model and hypotheses in \cite{d5}. The only
      possible ( if the phenomenon exists ) number  of flavours here is
      $n=6$. But the weak neutrality of $Sp(n/2)$  and immediate
      distinguishing between up and down leptons takes away  a picture
      for charged lepton spectrum.

The present paper is the continuation of paper \cite{d5}. We will
      demonstrate how the participation of $\nu$  in weak processes
      practically completely determines the structure of electroweak
      interactions with all leptons and quarks. Charged particles are
      present as observers, assuming the properties of $\nu$ and their
      spectrum.

      This is related with axial anomalies generated by the
      spectrum of six Dirac neutrinos.

      Let us explain this. Since the
      $\nu$ mass formation mechanism leads on its own to parity
violation, it appears redundant to introduce $R,L-$symmetry violation
by direct selection of specific (chiral) electroweak currents (weak
isospin $T_W$ and hypercharge $Y$). Let us assume that the currents are
vector and conserve parity: for all $R$ and $L$ components of leptons
and quarks, -- $T_W=1/2$; for $R$ and $L$ leptons, -- $Y=-1$; and for
all quarks, $Y=1/3$.

Expressions for neutrino contributions to these
currents upon their transformation into quasi-Majorana and then into
massive Dirac particles primarily result in tiny violations of $\nu$
      interaction universality $(\sim\mu^2/M^2)$ and nonconservation
      of lepton numbers $(\sim\mu/M)$.
      These phenomena disappear in the
      low-energy region, upon elimination of heavy $\nu$.  After that,
      however, contributions of light neutrinos become axial, i.e.,
      they lose parity and the axial anomaly appears to be inside the
      dynamic system under consideration.

      Only the electromagnetic
      current (for all light particles) and the left current of the
      weak isospin (compulsorily for both light leptons and quarks,
       neutral and charged components) do not have axial anomalies. For
       the neutral weak current itself, we obtain the known $Z$ boson
       current with the Weinberg angle \cite{d1}. Only these currents do
       not depend on large masses and remain acceptable for the
       theoretically consistent low-energy system of leptons and
       quarks.

       Chiral anomalies of low-energy currents also restrict
       the number of quark doublets and the number of light charged
      leptons: they cannot exceed the number of light neutrinos, which
      is three. The anomalies also prohibit mixing of light and heavy
      neutrinos.

         Currents without anomalies are the SM currents.

       Although it reproduces and explains many of the well-known
      properties, this approach does not provide a thorough insight
      into the structure of the SM electroweak part.  First, it does
      not penetrate into the Higgs part of the system, i.e., does not
      allow understanding of the mechanism for formation of the masses
      $M_W$, $M_Z$, or masses of charged fermions. The Higgs mechanism,
      though, offers no ultimate solution to this problem: it cannot
      explain the observed, clear hierarchy of charged fermion masses
      or the hierarchy of quark mixing angles \cite{d2}. The gauge group
      $Sp(3)$ demonstrates, however, that Higgs scalar states cannot
      arise from fundamental (present in the initial Langrangian)
      fields. At the same time, the spontaneous generation of Dirac
      masses can result in the appearance of scalars in the channel
      $\bar \psi_R\psi_L$, as it happens in the Nambu - Jona-Lasinio
model \cite{d7}.

     Secondly, the nonperturbative problem remains unsolved: what
     happens if the large scale $M$ increases in a system where both
     "good" (not leading to an anomaly at $M\rightarrow\infty$) and
     "bad"  (anomalous at $M\rightarrow\infty$) currents are present?
     What influence and how much involvement would heavy particles have
     in these "bad"  currents? In the case of neutrino weak currents,
     the "anomalous parameter" that is usually considered \cite{d9} is
     exceptionally great: $>10^{15}$.

     Using the terminology of the SM
     where any number of generations is possible, we can formulate a
     clear analog to this problem: what happens to three "good"
      generations if the mass $t'$ of the fourth generation quark (or
      the respective Yukava constant) increases and if at $m_{t'}
      \rightarrow\infty$
      only this generation produces a chiral anomaly?

      There is no
      solution to this problem.  Indirect arguments lead one to believe
      that anomalous currents will drop out, whereas the three "good"
      generations will reproduce the SM.

      In section 2, we briefly
      explain the logic of paper \cite{d5} and present the formulas
      necessary for the purposes of this paper. In section 3, neutrino
      contributions to neutral vector currents are expressed through
      mass states $\nu$. In section 4, all properties of SM electroweak
      currents are derived using the gauge $Sp(n/2)$ approach. Section
      5 presents the main results and implications of the proposed
      scheme.

\section {Gauge mechanism for neutrino mass generation}

      Let us retrace the logic of paper \cite{d5} and reproduce the
      formulas necessary for the purposes of the present paper.

Let us assume that $n$ lepton flavors are related with each other by
    means   of gauge transformations. This means that interactions which
      involve leptons (all $\nu$ and $e$ flavors
      \footnote{Separation of $\nu$ from charged $e$ flavors
      immediately comes from the quasiMajorana states choosing, see
      eq.(4).}) are invariant (locally or globally - is of no significance at this
      point) to transformations of some group. The symplectic group
      $Sp(n/2)$ is distinguished because its invariant Majorana masses
      are identically equal to zero under the fundamental, spinor
      representation. Their expressions, through the commonly used
      chiral operators $\psi_{R,L}(x)=1/2(1\pm\gamma_5)\psi(x)$,
       are ($a=1,2,\ldots,n$
      operators for massless particles):
\begin{equation}
\bar \psi_R^ah_{ab}C\bar
\psi_R^{Tb}=\psi_{Ra}^TCh^{ab}\psi_{Rb} \equiv 0\,,
\end{equation}
 In eq. (2) we omitted the
       argument $x$ and will omit it further on.  The skew-symmetric
       matrix $h(n\times n)$:
\begin{equation} h^T=-h\,,\qquad
h^{ab}=-h_{ab}\,,\qquad hh^+=1\,,
      \end{equation}
-- relates the equivalent conjugate representations $\psi_a$ and
$\psi^{+Ta}$.  It is a matrix of alternate numbers $\pm 1$ on the
right diagonal \cite{d10}. The matrix of the charge conjugation
$C$ has common properties: $C = -C^T$, $CC^+ = 1$.

In
$Sp(n/2)$, covariant analogs of Majorana states can be written in the
form:
\begin{equation}
\Psi_{(R,L)a}=\psi_{(R,L)a}+\gamma_5h_{ab}C\bar \psi_{(R,L)}^{Tb}\,,
\end{equation}
$\gamma_5$ are present in (4) due to the antisymmetry of $h$, eq.(3),
and are absent in representations and groups with $h = h^T$. One can
easily check the fulfillment of "Majorana conditions":
\begin{equation}
\Psi_R=\gamma_5hC\bar \Psi_R^T\,,\qquad \Psi_L=-\gamma_5hC\bar
\Psi_L^T\,.
\end{equation}
 The phase
for $\Psi_L$ in eq.  (5) is selected such ($-\gamma_5$ is introduced
for $L$ in (4)) that the Dirac part of the mass matrix
      ($\bar \psi_R\psi_L+\bar \psi_L\psi_R$
members) becomes real (see \cite{d5}).

      It is rather quasi-Majorana
combinations (4) than chiral states that play a major role in neutrino
mass formation dynamics.  Apparently, $\Psi_R$ and $\Psi_L$ are not
      entirely Majorana states (particle $a$ + antiparticle $n+1-
      a$):  they are four-component ones, but there are common, similar
      states in flavor-different $\Psi_{(R,L)a}$
      so that the entire
      number of independent states in both $\psi_{(R,L)a}$ and
      $\Psi_{(R,L)a}$, $a=1,2,\ldots,n$, ends up being the same.
            Such a
      mechanism, obviously, sets simultaneously apart "neutrino" states
      from equivalent components of the weak isospin in the original
      Lagrangian, which is invariant to both $Sp(n/2)$ and weak
      interactions.

      In terms of (4), operator members corresponding to
      the Majorana and Dirac parts of the mass matrix $\nu$ can be
written in the following form:  \begin{equation} \bar
\Psi_R^a\Psi_{Rb}\,,\quad \bar \Psi_L^a\Psi_{Lb}\,;\qquad \bar
\Psi_R^a\Psi_{Lb}\,,\quad\bar \Psi_L^a\Psi_{Rb}\,.  \end{equation} In
       \cite{d5}, it is shown how chiral representations of $Sp(n/2)$
       currents (and others ones) are related with Majorana
       representations through operators (4). Up to this moment, using
$\Psi_{(R,L)}$ instead of $\psi_{(R,L)}$ has simply been an equivalent
change of variables.

      This situation changes
when neutrino masses are assumed to result from the spontaneous
violation of the $Sp(n/2)$ symmetry, i.e., the vacuum averages of
operator combinations (6) (at least part of them) become numbers other
than zero. Using these vacuum averages, the complete mass matrix for
      neutrinos has the form $(2n\times 2n)$:
\begin{equation}
M=\left|\begin{array}{ll}
M_{RR}& \mu_{RL}\\
\mu_{LR}& M_{LL}\end{array}\right|\,.
\end{equation}
 Let us consider $M$ as a
      symmetric, real matrix ($CP$ is preserved).  Owing to Majorana
      conditions (5), the elements of these matrices are related with
      one another:
\begin{equation}
M^{+T}=-h^+Mh\,,\qquad \mu^{+T}=+h^+\mu h\,.
\end{equation}
 Relations (8) result in the characteristic
equation for $M$ roots being only dependent on eigenvalue squares,
i.e., the $M$ spectrum is $\pm\lambda_i$.

      The existence of relations
(8) greatly hinders the occurrence of symmetry breaking. One requires
such a system of equations for the M parameters that would set these
parameters unambiguously.
Owing to (8), the number of "gap"  equations
      \cite{d7}, defining the parameters, exceeds the number of
      independent variables (elements of the matrix diagonalizing $M$
      plus the number of eigenvalues).

      As shown in \cite{d5}, it is the
      transition to the Majorana form (4) that leads (in any
      $Sp(n/2)$-invariant model) to an additional global symmetry of
      "gap" equations which turns part of these equations into
       identities. There is no other choice but to select a form of
       $M$, (7), that will meet matched conditions:  the number of
       equations less the number of independent parameters is equal to
       the number of identities among the equations.

       To solve the
       problem unambiguously requires that (physically interesting
       variant):

       a) the number of flavors $n=6$,

       b) the matrix $M_{LL}=0$, i.e., all its elements be equal to
       zero ($L\rightleftharpoons R$ is certainly possible),

       c) $\mu_{LR}$ be
       a diagonal matrix with similar eigenvalues.

      Condition (c) should, in fact, be realized automatically,
       as the invariant, Dirac part of the matrix $M$ does not vanish
       in $Sp(n/2)$. One would believe that the spontaneous violation
       should follow here exactly this way of the least symmetry
       breaking.

The first two conditions, their purpose and meaning,
      are discussed in detail under Introduction. The possible action
      of the see-saw mechanism is confirmed by condition (b). The
      assumption of a great difference between the Majorana scale $M$
      and the Dirac $\mu$ resulting in the remarkable number of flavors
      $\mu$  three light and three heavy - as well as the specific
      character of matrices $M_{RR}$ (8) facilitate reproduction of the
      known properties of the $\nu$-spectrum (1) and lead to the
      coincidence with the observed number of light neutrinos.

      Diagonalization of the mass matrix meeting conditions (a, b, c)
      and (8) results in the eigenfunctions of matrix (7) (see
      \cite{d5}, eq.(36)):
\begin{equation}
\Psi_{\pm D}=U_{\pm D}\,^a(\cos\Theta_{\pm D}\Psi_{Ra}+\sin
\Theta_{\pm D}\Psi_{La})\,,
\end{equation}
$$
\psi_{\pm D}=U_{\pm D}\,^a(-\sin\Theta_{\pm D}\Psi_{Ra}+\cos
\Theta_{\pm D}\Psi_{La})\,,$$
$D = 1, 2, 3$. Here, $U$ is the orthogonal
      matrix diagonalizing $M_{RR}(6\times 6)$. $\Psi_D$ corresponds to
the heavy masses $\nu(M_D\sim M)$ and $\psi_D$ represents the light
particles $(m_D=\mu^2/M_D)$.  Values of $\cos\Theta_D$,
and $\sin\Theta_D$ are:
\begin{equation}
\cos\Theta_D=\frac{1}{\sqrt{(1+(\mu/M_D)^2}}\,,\qquad
\sin\Theta_D=\frac{\mu/M_D}{\sqrt{1+(\mu/M_D)^2}}\,,
\end{equation}
$$\cos\Theta_{-D}=\cos\Theta_D\,,\qquad
\sin\Theta_{-D}=-\sin\Theta_D\,.$$ $\Psi_{Ra}$ and $\Psi_{La}$ are
massless "Majorana" states (4), in all 24.  Massive eigenstates of
matrix (7) $\Psi_D$ and $\psi_D$, also 24 in number, have properties
similar to (5):  \begin{equation} \gamma_5hC\bar
\Psi^{TD}=\Psi_{-D}\,,\qquad \gamma_5hC\bar \psi^{TD}=-\psi_{-D}\,,
\end{equation}
$h_{D'D}\equiv h_{-DD}=-h^{-DD}$.
Equations (11) relate states with opposite mass signs.  Equations
(9)-(11) will help solve the principal problem in the next section:
expressing massless Majorana states $\Psi_{(R,L)a}$ (or chiral
$\psi_{(R,L)a}$)
through functions of massive Dirac neutrinos. Further, in section 4, we
will express electroweak currents in terms of physical, massive (Dirac)
$\nu$.

\section{ Representation of chiral vector currents through Dirac
massive neutrinos}

      To solve this problem, let us determine Dirac states
by combining two quasi-Majorana spinors (9) with masses equal in
      absolute value into one entity.  This can be done in a standard
      way. First, let us build true Majorana massive states (all masses
      are positive) for heavy neutrinos:
\begin{equation}
\Psi_D^{(1)}=\frac{\Psi_D+C\bar \Psi^{DT}}{\sqrt{2}}=
\frac{\Psi_D+\gamma_5h\Psi_{-D}}{\sqrt{2}}\,,
\end{equation}
$$\Psi_D^{(2)}=-\gamma_5h\frac{\Psi_{-D}+C\bar \Psi^{-DT}}{\sqrt{2}i}
=\frac{\Psi_D-\gamma_5h\Psi_{-D}}{\sqrt{2}i}\,,$$
and for light neutrinos:
\begin{equation}
\chi_D^{(1)}=\frac{\psi_D+C\bar \psi^{DT}}{\sqrt{2}}=
\frac{\psi_D-\gamma_5h\psi_{-D}}{\sqrt{2}}\,,
\end{equation}
$$\chi_D^{(2)}=\gamma_5h\frac{\psi_{-D}+C\bar \psi^{-DT}}{\sqrt{2}i}
=\frac{\psi_D+\gamma_5h\psi_{-D}}{\sqrt{2}i}\,,$$
The signs are chosen for convenience. The second equations in (12) and (13)
      are obtained by using relations (11). Equations (12) and (13)
      are the first $Sp(n/2)$-noncovariant formulas, in which
      $h\equiv h^{D-D}=\pm 1$.

       Dirac states can be chosen by different, but
      physically equivalent methods. The difference lies in the
      definition of what is considered a particle and what an
      antiparticle in systems of heavy and light states (or, which is
      the same, in the definition of $\psi$ and $\bar \psi$ for Dirac
particles).

The obvious definitions of Dirac states (for heavy and
light $\nu$, respectively):
\begin{equation}
\Psi_{M_D}=\frac{\Psi_D^{(1)}+i\Psi_D^{(2)}}{\sqrt{2}}\equiv\Psi_D\,,
\qquad
\psi_{\mu_D}=\frac{\chi_D^{(1)}+i\chi_D^{(2)}}{\sqrt{2}}\equiv\psi_D\,,
\end{equation}
 after inverting expressions (9), result
in the following formulas:
\begin{eqnarray}
\Psi_{La}=U_a^{+\,-D}(-\gamma_5h^+\Psi_{M_D}^C\sin\Theta_D+\cos
\Theta_D\psi_{\mu_D})+U_a^{+D}(\Psi_{M_D}\sin\Theta_D-\cos\Theta_D
\gamma_5h^+\psi_{\mu_D}^C), \\ \nonumber
\Psi_{Ra}=U_a^{+\,-D}(\gamma_5h^+\Psi_{M_D}^C\cos\Theta_D+\sin
\Theta_D\psi_{\mu_D})+U_a^{+D}(\Psi_{M_D}\cos\Theta_D+\sin\Theta_D
\gamma_5h^+\psi_{\mu_D}^C)\,.
\end{eqnarray}
 To make eq.(15) covariant, the symbol
$h^+$ is introduced which corresponds to $h$ in (12)-(13). One should
mention, however, that all $h$ and $h^+$ are $(\pm 1)$ and selecting
either value makes no essential difference. $\Psi^C$ and $\psi^C$  are
antiparticle operators: $\psi^C=C\bar \psi^T$.

      The same symbol $D$
stands for both large and small masses. In the see-saw model, it
is reasonable to use the same symbol for masses with different
sign; this is how pairs of large-small masses are made up here
\cite{d4}. Then, the
      corresponding Dirac particles should be determined by the
      formulas:
\begin{equation}
\Psi_{M_D}=\frac{\Psi_D^{(1)}-i\Psi_D^{(2)}}{\sqrt{2}}\equiv C\bar
\Psi_D^T\,,\qquad \psi_{\mu_D}=\frac{\chi_D^{(1)}+i\chi_D^{(2)}}
{\sqrt{2}}\equiv\psi_D\,.
\end{equation}
In eq.(16) we retained the small mass as a
"particle". Relation (11) is used again in the first expression of
(16).

      Then, $\Psi_{La}$ and $\Psi_{Ra}$ are equal to:
\begin{eqnarray}
\Psi_{La}=U_a^{+\,-D}(-\gamma_5h^+\Psi_{M_D}\sin\Theta_D+\cos
\Theta_D\psi_{\mu_D})+U_a^{+D}(\Psi_{M_D}^C\sin\Theta_D-\cos\Theta_D
\gamma_5h^+\psi_{\mu_D}^C)\,, \\ \nonumber
\Psi_{Ra}=U_a^{+\,-D}(\gamma_5h^+\Psi_{M_D}\cos\Theta_D+\sin
\Theta_D\psi_{\mu_D})+U_a^{+D}(\Psi_{M_D}^C\cos\Theta_D+\sin\Theta_D
\gamma_5h^+\psi_{\mu_D}^C)\,.
\end{eqnarray}
 Formulas
      (15) and (17) allow expressing $Sp(n/2)$-invariant chiral vector
      currents through massive states. At first, by means of direct
      substitution we obtain:
\begin{equation}
\bar \psi_L^a\gamma_\rho\psi_{La}=-\frac 12 \bar\Psi_L^a\gamma_\rho
\gamma_5\Psi_{La}\,,\qquad
\bar \psi_R^a\gamma_\rho\psi_{Ra}=+\frac 12\bar \Psi_R^a\gamma_\rho
\gamma_5\Psi_{Ra}\,.
\end{equation}
Note that the Majorana vector currents:
\begin{equation}
\bar \Psi_L^a\gamma_\rho\Psi_{La}=\bar \Psi_R^a\gamma_\rho\Psi_{Ra}
\equiv 0
\end{equation}
-- are absolute neutrality of Majorana multiplets.

Transforming eq.(18) by using (15) and (17), we obtain:
\begin{eqnarray}
\bar \psi_L^a\gamma_\rho\psi_{La} &=&-\sin^2\Theta_D\bar
\Psi_{M_D} \gamma_\rho\gamma_5\Psi_{M_D}\,-\\ \nonumber
 &&\cos^2\Theta_D\bar\psi_{\mu_D}
 \gamma_\rho\gamma_5\psi_{\mu_D}-R_\rho(\Psi_{M_D},\psi_{\mu_D})
\frac 12\sin 2\Theta_D~,\\ \nonumber
\bar\psi_R^a\gamma_\rho\psi_{Ra} &=&\cos^2\Theta_D\bar \Psi_{M_D}
\gamma_\rho\gamma_5\Psi_{M_D}\,+\\ \nonumber &&\sin^2\Theta_D\bar
\psi_{\mu_D}
\gamma_\rho\gamma_5\psi_{\mu_D}-R_\rho(\Psi_{M_D},\psi_{\mu_D})
\frac 12\sin 2\Theta_D~,
\end{eqnarray}
where, for (15) and (17), respectively, the vector $R_\rho$ is
equal to:
\begin{equation}
R_\rho(\Psi_{M_D},\psi_{\mu_D})=\left\{\begin{array}{l}
\bar \Psi_{M_D}\gamma_\rho\psi_{\mu_D}^C+\bar \psi_{\mu_D}^C
\gamma_\rho\Psi_{M_D}\\
\bar \Psi_{M_D}\gamma_\rho\gamma_5\psi_{\mu_D} +\bar \psi_{\mu_D}
\gamma_\rho\gamma_5\Psi_{M_D}\end{array}\right.~.
\end{equation}
 Certainly, in formulas (20) and (21), summation over $D=1,2,3$ is
implied. To obtain these expressions, we used the orthogonality of the
matrices $U$.

The axial current has a simple formula, which is similar
for both cases (15) and (17):
\begin{equation}
\bar \psi^a\gamma_\rho\gamma_5\psi_a=\bar \Psi_{M_D}\gamma_\rho
\gamma_5\Psi_{M_D}+\bar \psi_{\mu_D}\gamma_\rho\gamma_5\psi_{\mu_D}
\,.
\end{equation}
 The vector current is:
\begin{equation}
\bar \psi\gamma_\rho\psi=\cos 2\Theta_D(\bar \Psi_{M_D}\gamma_\rho
\gamma_5\Psi_{M_D}-\bar \psi_{\mu_D}\gamma_\rho\gamma_5\psi_{\mu_D})-
\sin 2\Theta_DR_\rho(\Psi_{M_D},\psi_{\mu_D})\,.
\end{equation}
 Let us
      note the two properties of vector currents:

1. With regard to
heavy particles $(\sim M)$, transitions between light and heavy
neutrinos are possible, with both conservation and nonconservation of
the lepton number. This effect is negligible at
       $\mu<< M$ $(\sim\mu/M)$.

       2. The
presence in (23) of $\cos2\Theta_D$ attests to a small (at $\mu<< M)$
nonuniversality of neutrino vector interactions $(\sim\mu^2/M^2)$.

Both formulas (22) and (23) demonstrate that the transition to massive
states in currents (i.e., neutral currents, the subject of the next
      section) does not produce transitions between the types of light
      neutrinos (between flavors).

In conclusion, let us point out
      that the sign of the $\sim\sin 2\Theta_D$ contribution in eq. (23) depends on
       selection of $h\Rightarrow\pm 1$.

\section{ Electroweak currents}

During the spontaneous transition to massive neutrinos, $R,L-$symmetry
is not preserved. Let us show that this is a good enough reason for
distinguishing the observed currents of electroweak interaction,
namely, currents of $W-$bosons and the photon, from common vector
(nonchiral) currents which lack parity violation.

     In the SM, the
       interaction of W only with left chiral currents is a postulate:
       all left fermion components are weak isospin doublets and all
       right ones are singlets. In addition to weak isospin $T_W$
       currents in the SM, there is the $R,L-$asymmetric current of the
       hypercharge $Y$: $Y = 1/3$ for $L$ quarks, $Y=-1$ for $L$
       leptons; for $R$ quarks $Y = 4/3$ for $u$ and $Y=-2/3$
       for $d$ $R$ quarks; $Y=-2$ for charged $R$ leptons, and $Y=0$
       for $R$ neutrinos \cite{d1}. As a result, we obtain the
       electromagnetic and left weak currents, which explain the
       observed phenomena.

Let us take, as the basis for the
       electroweak theory in this paper, $R,L-$symmetrical (i.e.,
       preserving parity) full vector currents of all initially
       massless leptons and quarks. In such currents, the
characteristics of both right and left particles should be
similar. Therefore, not only $L$ but also $R$ components are
doublets of the weak isospin; all $L$ and $R$ quarks have $Y=1/3$;
and all $L$ and $R$ leptons have $Y=-1$:
\begin{equation}
{\rm J}_{W\rho}=g\sum_{f}(\bar \ell\gamma_\rho{\rm T}_W\ell+\bar
q\gamma_\rho T_Wq)\,,
\qquad
 J_{Y_\rho}=g'\sum_{f}\left(\bar \ell
\gamma_\rho\frac Y2 \ell+\bar q\gamma_\rho\frac Y2 q\right)\,.
\end{equation}
 All fermion
operators have a four-component Dirac form, and the sum is carried out
over all types of particles (including color for quarks). The lepton
part of currents is also invariant to $Sp(3)$, whereas quarks can be
neutral towards this group.

Then, the fundamental theory for massless
     particles, invariant to $Sp(3)$, $SU_W(2)$, $SU(3)$ (color) and
     $Y$, has only vector currents, lacks anomalies, and is entirely
     defined and renormalizable. Of course, it should be supplemented
     with the mass formation mechanism for $W$, $Z$, and charged
     leptons and quarks.  Let us assume that this mechanism involves
     violation only of the weak $T_W$ group and that no new $R,L-$
     symmetry violations occur.

     Compulsory participation of Majorana
     states in the $Sp(3)$-group spontaneous violation model under
     discussion requires some additional explanations in connection
     with  the appearance of the quantum numbers $T_W$ and $Y$.  As is
     known, $\psi$ and $\bar \psi$  have different values of $T_3$ and
     $Y$ and it appears impossible to combine them into a Majorana
     object.  We have earlier explained, however, that the dynamically
     active states (4) being used are quasi-Majorana ones, and their
     $\psi$ and $\bar \psi$  have different characteristics $(Sp(n/2)$
     representation components, $a$ and $n+1-a$). The states of
     this kind resemble Bogolyubov \cite{d8} systems: a particle with
     one spin projection + "hole" with the opposite spin projection.
     The above complexity, however, is of purely technical
     significance:  what should be the form of state (4) (for example,
     a doublet). As such, it has no implications for our discussion of
     vector currents; the results of this paper and paper \cite{d5} are
     not affected.

Let us express the neutrino contribution in eq.(24)
     through Dirac mass states, i.e., substitute in .(24) expression
     (23) and neglect contributions from heavy masses. For neutral
     currents, we obtain:
\begin{equation}
-\bar \psi_{\mu_D}\gamma_\rho\gamma_5\left({T_3\atop Y/2}\right)
\psi_{\mu_D}+
\{\mbox{vector part}\, T_3\, \mbox{or}\,  Y/2 \,
\mbox{},
\end{equation}
$$L+R \,\, \mbox{components of charged leptons and}\, u \,
\mbox{and}\, d\, \mbox{quarks}\}~.$$

      So, the
low-energy part appears to be axial and contains anomalies induced
by contributions from interactions between vector bosons of the
theory presented by the well-known triangle diagrams \cite{d1} .

     Only those anomalies are important that are
     caused by weak and hypercharge vector bosons: $W^\pm$, $W_0$ and
     $B$.  Color currents of quarks apparently do not introduce
     anomalies as they have no relation to neutrinos. The gauge bosons
     $Sp(3)$,  $F_\mu$  in terms of \cite{d5}, should acquire heavier
     masses if such a complex spontaneous violation of the group
     (involving all its operators) does take place. At the energy of
     the order of these masses , no anomalies are present due to the
     full vectorness of the whole interaction system.

     There are six
     possible combinations of weak $W$ and hypercharge bosons $B$ in
     which an anomaly can occur. These combinations are calculated and
     sorted by $R$ and $L$, for leptons and quarks. We do not  cancell
     quark and charged lepton contributions due to their vectorness,
     i.e., in the sum of $L+R$ contributions:  as the left world is
     now different from the right, it is important to know the state of
     things in each of them separately. The following table shows the
     results of the calculations.

          {\bf Low-energy anomalies by $R,L$ sectors; leptons $(\ell)$
and quarks $(q)$}

\vspace{0.5cm}

\begin{center}
\begin{tabular}{|c|c|c|c|c|} \hline
&\multicolumn{2}{c|}{R}&
\multicolumn{2}{c|}{L}\\ \hline
& $\ell$ & $q$& $\ell$& $q$\\ \hline
$BBB$& 0 & +2/9& +2 & -2/9\\ \hline
$W_3BB$ & -2& 0& 0& 0\\ \hline
$BW^+W^-$& 0& 2$\Theta_{qR}^2$& 2$\Theta_\ell^2$& -2$\Theta_{qL}^2$
\\ \hline
$W_3W^+W^-$& 0& 0& 0& 0\\ \hline
$BW_3^2$& 0& +2& +2&-2\\ \hline
$W_3W_3^2$& -2& 0& 0& 0\\ \hline
\end{tabular}
\end{center}

\vspace{0.5cm}
\noindent

The calculated numbers are coefficients at anomalous
divergencies\cite{d11}:
\begin{equation}
\partial_\mu j_\mu^{(5)}=\frac{g_1g_2g_3}{(4\pi)^2}\; \mbox{Tr} F\bar F\,.
\end{equation}
 They take into account signs and values of the $Y(Y_\ell =
-1, Y_q = 1/3)$ and $2T_3$ components $(\pm 1)$ as well as the
difference in sign of chiral $R(+1)$ and $L(-1)$ contributions.

When
calculating table contributions from neutral currents, we used the
     obvious formulas of the low-energy expression (25) for one
     generation (for light masses $\mu_D$ for $\nu$ and $m_d$ for
     $\ell$):
\begin{equation}
J^{(T_3)}=\bar \nu_L\gamma\nu_L-\bar \ell_L
\gamma\ell_L-\bar \nu_R\gamma\nu_R-\bar \ell_
R\gamma\ell_R+\bar u_i\gamma u_i-\bar d_i\gamma d_i\,,
\end{equation}
$$
J^{(Y)}=-\bar \nu_L\gamma\nu_L-\bar \ell_L
\gamma\ell_L+\bar \nu_R\gamma\nu_R-\bar \ell_
R\gamma\ell_R+\frac 13 \bar u_i\gamma u_i+\frac 13
\bar d_i\gamma d_i\,,
$$
 summed over the color i
     In eq.(27), the charges $g/2$ and
     $g^1/2$, respectively, are omitted.
Sign difference of the $R$ and
     $L$ parts results from axiality of the neutrino contribution in
eq.(25).

Investigating charged current participation in $BW^+W^-$ and
$W_3W^+W^-$ is more complicated, although more informative. At energies
$E << M$, there is no interaction of light neutrinos from eqs. (15),
(17) with right charged leptons ("electrons"), because large
contributions in (15), (17) for $R$ neutrinos, $\nu_R = (1/2)(1 +
\gamma_5)\Psi_R$, include only heavy particles $(M_D)$.
     Therefore, at $E<<M$ charged currents contain only
$\ell_L$ components:
\begin{equation}
\bar \ell_L^a\gamma\nu_{La}+\bar \ell_R^a
\gamma\nu_{Ra}+c.c.\simeq \bar\ell_L^a\gamma\frac 12
(1-\gamma_5)\left(U_a^{+-D}\nu_{\mu_D}-U_a^{+D}\gamma_5C
\bar \nu_{\mu_D}^T\right)+c.c.
\end{equation}
 The four-component Dirac spinor of the massless electron $\ell$
     should be expressed through massive states of all electron flavors
     (at that, the origin of $\ell$ masses is of no importance).
     Generally, we have:
\begin{equation}
\ell_a=A_a^d\ell_{m_d}+\widetilde{A}_a^d\ell_{M_d}\,,
\end{equation}
$A_a^d$ ¨ $\widetilde{A}_a^d$  $(6\times 3)$.
 Where $m_D$ and $M_d$ are masses of light and
     heavy electrons and $A_a^d$ and $\widetilde{A}_a^d$
 are some matrices $(6\times 3)$.  We
     have assumed that the number of light electrons is equal to the
     number of light neutrinos. An inequality would lead to the lack of
nonanomal currents in a low-energy system, i.e., to the absence of a
low-energy limit independent of heavy masses ( see the discussion of
eq.(37) ).

The superpositions
\begin{equation}
\ell_{LD}^{(1)}=U_{-D}\,^aA_a\,^d\ell_{m_dL}\quad \mbox{and}\quad
\ell_{LD}^{(2)}=U_D\,^aA_a\,^d\ell_{m_dL}
\end{equation}
 appear in eq.(28) as
representatives of $L-$electron states.
The interaction of $\ell_{LD}$ with $R$
and $L$ neutrinos and antineutrinos does not conserve the lepton number
\begin{equation}
\bar \ell\gamma\nu+c.c.=\bar \ell_{LD}^
{(1)}\gamma\nu_{\mu_DL}+\bar \ell_{LD}^{(2)}\gamma
C\bar \nu_{\mu_DR}^T+c.c.
\end{equation}
The interaction $W^\pm$ with $\ell_{LD}^{(2)}$ results in four
additional anomalous contributions \footnote{The propagator of the
antiparticle $\nu^C$ assings a sing opposite to normal $R,L$ cases
to the liner divargence of the diagrams \cite{d11} and
consequently to the anomaly: $<C\bar \nu_R^T,\nu_R^TC>=-C<\nu_R^T,
\bar \nu_R^T>C\rightarrow-C\hat p^TC=-\hat p$.},
     which are not included in the Table. These
contributions are proportional to quantities that can be termed as
generalized sums of squares of lepton mixing angle cosines.  In the
presence of both members (31), we have two sums over all low-energy
lepton flavors (refer to a single generation):
\begin{equation}
\Theta_{\ell L}^2=\frac 13\sum_{D,d}A_d^{+a}U_a^{+-D}U_{-D}\,^{a'}
A_{a'}^d\,, \qquad \Theta_{\ell R}^2=\frac
13\sum_{D,d}A_d^{+a}U_a^{+D}U_{D}\,^{a'} A_{a'}^d\,.
     \end{equation}
     At
$\Theta_{\ell R}^2\neq\Theta_{\ell L}^2$ (using Table designation
$\Theta_{\ell L}\equiv\Theta_\ell^2$)) but at $\Theta_{\ell R}^2\neq
     0$, currents without anomalies are not present at all and there is
no low-energy limit independent of  large masses.  At $\Theta_{\ell
     R}^2=\Theta_{\ell L}^2$ (for which there is no apparent reason),
there is one non-anomalous   electromagnetic   current (see eq. (35)),
whereas $R$ and $L$ weak $T^\pm,T_3$ currents are anomalous.

The only
scenario that results in a complete spectrum of independent low-energy
currents is the exclusion of the interaction that violates the lepton
number.In this scenario, the matrix $A$ should have the form:
\begin{equation}
A_a^d=U_a^{+-D}V_D^d\rightarrow\Theta_{\ell R}^2=0\,,
\end{equation}
since for the unitary matrix $U$ we have $U_D\,^aU_a^{+-D}=0$.
Then
\begin{equation}
\Theta_\ell^2=\Theta_{\ell L}^2=\frac 13 \sum_{d,D}V_D^{+d}V_d^D=
\frac 13 \sum_{D,d}\cos^2\Theta_{dD}\,,
\end{equation}
$V_D^d$ is a matrix and $\Theta_{dD}$ are mixing angles of light
     leptons.  The Table shows this very pattern of anomaly
     distribution. The quantity
     $\Theta_{qL}^2=\Theta_{qR}^2\equiv\Theta_q^2$
          for quark generations
is similar to (34) for leptons.

     Then, based on the Table, the left
current $T_{3L}$ (the interaction with $W_3$) is apparently
non-anomalous.  At the same time, the electromagnetic current
interacting with the photon\cite{d1}
\begin{equation}
j_{e m}=e Q=e\left(T_3+\frac Y2\right)~,\qquad e=\frac{g
g'}{\bar{g}}~,\qquad\bar{g}=\sqrt{g^2+g^{'2}}
\end{equation}
 is also non-anomalous provided that
($T_3$ and $Y/2$ refer to both $L$ and $R$ components):

a) the
quantity $\Theta^2_{qR}$ can be excluded from consideration

b) $\Theta^2_\ell =\Theta_q^2$.

Right lepton contributions to charged current $T^\pm$ anomalies are
absent owing to the low-energy interaction structure (31) resulting
from (33):
\begin{equation}
\bar \ell\gamma\nu+c.c.=\bar \ell_{m_dL}V_d^{+D}\gamma\nu_{\mu_DL}+c.c.
\end{equation}
The contribution $\Theta_{qR}^2$ from right quark currents can be
neglected in the system of only left weak currents. Condition (a) means
selecting only left weak currents. The equality (b) of lepton and quark
expressions (34) independent of each other makes sense only in one
instance, namely, when systems of light leptons and quarks are
complete:
\begin{equation}
\Theta_\ell^2=\Theta_q^2=1\,.
\end{equation}
This means that there should be three flavors of light
electrons, there should be no mixing of light and heavy $L$ leptons,
and there should be only three generations of quarks. At $d = D = 3$,
the representation of eq. (33) for $A_a^D$ is already equivalent to the
absence of mixing.

Also note that eq.  (33) indicates a close
relationship between neutrino and electron spectra: $U$ is the matrix
diagonalizing the Majorana mass matrix.  This relationship is in
     conflict with the Higgs mechanism of mass formation where spectra
     of up and down components are independent, being determined by
     different $SU_L(2)$ invariants of Yukawa couplings.

     One can see
     from the Table that all anomalies from all $R$ and $L$ components
     of light particles in superposition (35) are reduced. The
     electromagnetic current (35) includes the currents $T_3$ and $Y/2$
of both $L$ and $R$ sectors.

     Consequently, the Table corresponds to a
system of non-anomalous currents consisting of three left weak currents
that interact with $W$ and the full electromagnetic current (35) that
     interacts with the photon. Only these currents are independent of
     the high-energy part of the scheme. Let us point out that the
     difference of currents other than $T_L$ and $Q$ is not that
     $T_R=0$, as it is believed in the SM, but that these currents are
     anomalous, i.e., require involvement of heavy particles. $W$ also
     interacts with $R$ components; at that, participation of heavy
     fermions is imperative, as in the instance of the charged current
     with $\ell_R$ participation (see eqs. (15) and (17)).

Therefore,
     the current interacting with the vector $Z-$boson \cite{d1}
     orthogonal to the photon:
\begin{equation}
\bar g(T_3-\sin^2\Theta_WQ)\,, \qquad\qquad \sin\Theta_W=\frac{g'}
{\bar g}
\end{equation}
 is non-anomalous and is
     consistently low-energy only if $T_3\equiv T_{3L}$, i.e., contains
     only left components of $\ell$ and $q$ massive states, as is
     postulated in the SM.

      The system of left weak and full
     electromagnetic currents is the very system of SM fermion
     currents. It appears to be the only anomaly-free system and
therefore is independent of high-energy physics in our scenario.
Our mechanis of the low-energy anomaly contraction for weak currents
 essentially differs from the same in SM. There we had independent
contraction for any row of the Table. Here the anomaly is absent only
for the sum of currents forming the observed electromagnetic one.

The analysis of charged currents $T^\pm$
proved to be heuristically important:

      1. The anomalous character of $R-$charged current has been
      proved:  taking it into consideration leads to the quark anomaly
      $\Theta^2_{qR}$ that is not compensated in $R$ components. This
does not permit turning the neutral electromagnetic current into
non-anomalous.

2. The $L-$quark anomaly is coupled with the $L-$lepton
anomaly and they compensate each other.  Therefore, although the
$R,L-$symmetry was violated only in the lepton sector, non-anomalous
electromagnetic current would not be possible  in the absence of
quarks.

3. It is charged currents that limit the number of quark
generations.

The SM allows any number of generations. In the proposed scenario,
attempts to change the number of quark generations and light
electrons (only three kinds of light neutrinos are possible, see
\cite{d5}) would lead to the disappearance of the entire system of
non-anomalous
      low-energy currents independent of high-energy physics. These
      currents exist only in three generations of particles included in
      the SM.

The consistent and decisive consideration of the problem
      would include investigating the change in the properties of a
       well defined system (without anomalies) with the increase in the
       large "anomalous" parameter $(M)$. This could shed light on the
       "fate"  of all currents and define more specifically the
       influence of heavy masses on currents that remain anomalous at
$M\rightarrow\infty$. We can only indicate that in the
       proposed scheme currents that are non-anomalous, and therefore
       the best-suited for independent low-energy survival, are the
       ones of the SM: the left weak and the total electromagnetic
       currents.

      As discussed in Introduction, our study is not
       complete. We only can  repeat here that the existence of
       doublets of both left and right components does not allow
      inclusion of the Higgs scalar field with required properties
      $(T_W = 1/2)$ in the fundamental Lagrangian with Yukawa
      couplings.  The immediate result is the violation of the weak
      $SU(2)$ symmetry.  At the same time,  scalar fields $T_W=0$ and
      $T_W = 1$ may participate in the dynamic of mass formation.

\section{ Conclusion}

      Notwithstanding the unexplained masses $W$, $Z$
      and charged leptons-quarks, the hypothesis of the gauge
      $Sp(3)$ symmetry of lepton flavors is remarkable in that it
      allows one to relate and explain the observed facts and proposed
      properties that form the basis of the SM, as well as offer
      answers to a number of fundamental questions.

      1. The number of
light neutrinos is three (plus three very heavy ones). The same applies
to charged leptons. Light leptons do not mix with heavy ones.

      2. There
may only be three generations of quarks (there is no reason to expect a
greater number of  generations).

      3. The $R-L$ asymmetry in spontaneous
neutrino mass formation is the sufficient cause of weak parity
nonconservation.

      4. The see-saw mechanism is necessary for the
appearance of $\nu$ masses.

5. There can be similar hypercharge values
and a common weak isotopic pattern for Dirac states of all leptons and
quarks.

In weak currents, before the mass $\nu$ formation mechanism steps
in, parity could conserve.

6. Electromagnetic, $Z$ bozon and weak
charged currents are distinguished at low energies.

7. A simple
explanation of neutrino spectrum features becomes possible.

8. The
double $\beta-$decay is absent in the dynamics under consideration.

    The author is grateful to Ya. I. Azimov , G. S. Danilov and V.Yu.Petrov for useful and enlightening discussions.

\end{document}